\documentclass [a4paper, cite, 11pt] {amsart}
\usepackage [latin1]{inputenc}
\usepackage[all]{xy}
\usepackage{amsmath}
\usepackage{amsfonts}
\usepackage{latexsym}  
\usepackage{amssymb}
\usepackage{graphics}
\usepackage{graphicx}
\newtheorem{teorema}{Theorem}[section]
\newtheorem{Lemma}[teorema]{Lemma}
\newtheorem{propos}[teorema]{Proposition}
\newtheorem{corol}[teorema]{Corollary}
\newtheorem{ex}{Example}[section]
\newtheorem{rem}{Remark}[section]
\newtheorem{defin}[teorema]{Definition}
\def\bt{\begin{teorema}}
\def\et{\end{teorema}}
\def\bp{\begin{propos}}
\def\ep{\end{propos}}
\def\bl{\begin{Lemma}}
\def\el{\end{Lemma}}
\def\bc{\begin{corol}}
\def\ec{\end{corol}}
\def\br{\begin{rem}\rm}
\def\er{\end{rem}}
\def\bex{\begin{ex}\rm}
\def\eex{\end{ex}}
\def\bd{\begin{defin}}
\def\ed{\end{defin}}

\def\C{{\mathbb C}}

\begin {document}
\title[Dr.\ Strangelove]{Dr.\ Strangelove or: How I Learned to Stop Worrying and Love the Citations}

\author[Alberto Saracco]{Alberto Saracco}
               \address{Dipartimento di Matematica e Informatica, Universit\`a di Parma, Parco Area delle Scienze 53/A, I-43124 Parma, Italy}\email{alberto.saracco@unipr.it}
                             
               \date{\today}

\begin{abstract} 
Citations are getting more and more important in the career of a researcher. But how to use them in the best possible way?

This is a satirical paper, showing a bad trend currently happening in citation trends, due to intensive use of citation metrics.

I am putting this on the arXiv and on Researchgate. Should you be interested to publish this paper on a journal of which you are editor, let me know.
\end{abstract}
 \maketitle

\section{What's citations got to do with it?}

In ancient times, citations in research papers were meant to allow readers to know two things: firstly the cited paper was somehow relevant to the research done in the paper, secondly the cited paper was a suggested read to the interested reader.

But in the last 10 to 15 years, citations are meant to do something totally different: to measure how good a researcher is. The more citation they get, the better the paper (or the researcher, or the group of research, or journal) is! 

Highly cited papers, individual and collective H-index, sheer number of citations, all theese became a clear and \textit{objective} signal of being a good researcher.

In Italy the number of citations comes in $2$ different necessary conditions in order to have a national habilitation (\textit{ASN --- Abilitazione Scientifica Nazionale}): at least $2$ out of $3$ magical values, sheer number of citations, $H$-index (and number of published papers\footnote{Everything must be certified by Scopus or Isi-Web-Of-Science, which is a great idea, since they are private companies on whom the academic community has no control}), must be met by a researcher who wants to apply to a habilitation to professorship. But Italy is just an example on how the \textit{objective} numbers of citations (and papers) have a central role in the researcher's fundings and career. While once Academia was about \textit{publish or perish}, now things got tougher, now Academia is about \textit{publish (possibly in highly ranked journals) and get cited (a lot) or perish}.

One might object that this use of citations is totally different from what they were meant. And after all \textit{what's citations got to do with it (the merit of the reasearcher)?}

\section{When the going gets tough...}

Of course the previous objection is totally silly: we cannot pretend to live in the past and decide by ourselves what citations are meant for.

We must totally change our point of view and accept the rules as they are and play according to the rules. After all rules are chosen so that people are encouraged to behave in a positive way. And also in this case it is working: more cites mean a better researcher? And indeed lately the number of citations is raising: always better researches, year after year.

Whenever we have rules and players, we are in the kingdom of game theory. Game theory always supposes the players to be rational players. And who can be more rational than a researcher? The rules and the rationality of researchers totally accounts for the rise in citations we are experiencing.

So the tough (and rational) researcher gets going by learning how to implement in the best possible way the citations in his or her papers. The optimum would be a paper with no text and citations to all of your previous works \cite{AS,ARS,AlS,BS,DS07,DS11,MS,NS,SbS,S08,S17,S18a,S18b,SS,SaT,ST07,ST08,ST11}, but somehow some old-schoolers make some resistance and do not allow such perfect papers to be published.

In this paper we want to show a method to go as near as possible to the optimum, keeping in mind that if your paper does not get published, you are getting zero citations immediately and zero citations afterward.

\section{All you need is love (and citations)}

Obviously, the most citations of your works, no matter if already published, if pre-print \cite{MS18,MS19,MS19b,Top} or future \cite{ASS,BSZ,BSS,BS,BST,GS,PSV,Qua}, you get to put in the bibliography, the better it is. And remember that also citations to pre-prints or future papers will work: you just have to remember to show the citation to Scopus or Web Of Science. Many Universities have people whose work is partly to raise the citations of the University researchers\footnote{Just think for a second how lovely useful this work is!}, so make sure you tell them about your future citations. 

So make sure you cite everything that you wrote or are writing or will write and is (even marginally) relevant to your paper!

Of course a series of good practices to maximize citations includes (but is not limited to) the salami slicing technique (divide a paper in several least publishable units) and putting a slight error in your paper you will have to correct with an errata later on.

\section{Spreading the love}

But remember: love (and citations) is about sharing, so don't be selfish.

Don't be shy and cite your own brother \cite{Gio1,Gio2,Gio3} or your younger collaborator in need of an habilitation \cite{Sam1,Sam2,Sam3}.

You can even cite papers recently published in your University's journal \cite{Par1,Par2,Par3}, in order to raise its cite score and ranking; or papers from your Department's colleagues \cite{mat1,mat2,mat3} in order to raise the Department's H-index. These two things are useful since these numbers are used, and possibly in the future will be used even more, to determine how good a Department is and how much fundings it will get. So, never forget to take an occasion to raise these numbers.

But also never forget that in order to get published, you have to convince at least one referee and one editor. So carefully choose your editor and cite him or her \cite{ed1,ed2,ed3}) and try to suggest a referee by citing him or her a lot (often referees are choosen between researchers cited in the paper) \cite{ref1,ref2,ref3}. Remember that citations are love and everybody likes to feel loved. An editor or referee who can feel your love is more likely to accept your paper.

Finally, cite some papers recently published in the journal you are submitting your paper to \cite{jou1,jou2,jou3}. This shows you truly believe that is an important journal. Moreover, if your paper gets published, the journal numbers will get better.

But of course not all citations are equal. The most effective citations are those which increment the H-index of the researcher. Hence a researcher should know how to cite in order to maximize the probability of getting to a certain H-index in a given amount of time. Thus, in this paper I have chosen to give citations to other people in order to try to maximize their H-indexes.

\section{Conclusions}
Obviously, applying the above suggested way of conduct to your paper, after adding the really-related citations to your paper, you might end up with a several pages long bibliography. But that, rather than being a problem, is an objective index of how much high-quality research has been published lately.

So don't ever be distracted by emotionality and by a fake sense of integrity or morality: you are a tough researcher and should behave accordingly to the rules, maximize the citations and spread the love.

Of course, never forget to cite this paper whenever possible. Spread the love!

\section*{Conflict of interests}
The author declares that this paper, if published, will drastically increase his citation numbers, having cited all of his past present and some of future works. This does not violate any ethics conduct as explained in the paper itself and it is moreover useful to exemplify the phenomenon we are studying.

\section*{Acknowledgments}
The idea beneath this paper was born in a Facebook discussion of the author with Daniele Castorina and Marco Isopi. I wish to thank them, of course by citing some of their papers \cite{Cas1,Cas2,Cas3,Iso1,Iso2,Iso3}. Let's spread the love!

\section*{Getting serious for a moment}

Getting serious for a moment, all of the above is obviously something that should be totally avoided. It obscures the really relevant citations in a sea of irrelevant ones. Yet no one can deny it happens. Researchers tend to self-cite or group-cite much more, often the referee asks to add some very important citations, usually all to articles by the same author.

No one can deny the medium number of citations per paper rose a lot in the last 10-15 years. Why? Of course for the reasons highlighted in this paper. Give a goal, and people will follow it.

When rules are decided, this should be kept in mind. It does not matter whether or not citations are a good tool to measure the goodness of a research. What it matters are the consequences of the rules. Rules should be chosen in order to have people behave in a better way. The ranking-mania with an extensive use of citations proved itself to have bad consequences. A better method has to be thought. 

In order to make citations matter again, we should stop using them to evaluate papers, journals and researchers.

\end{document}